\documentclass[%
 reprint,
 amsmath,amssymb,
 aps,
]{revtex4-1}

\usepackage{graphicx}% Include figure files
\usepackage{dcolumn}% Align table columns on decimal point
\usepackage{bm}% bold math
\usepackage{xcolor}

\begin{document}

\preprint{APS/123-QED}

\title{Quantifying the non-equilibrium activity of an active colloid}% Force line breaks with \\%

\author{Sarah Eldeen$^{1}$, Ryan Muoio$^{1}$, Paris Blaisdell-Pijuan$^{1,3}$, Ngoc La$^{1,4}$, Mauricio Gomez$^{1}$,  Alex Vidal$^{2}$,  and  Wylie Ahmed$^{1,\star}$ }

\affiliation{$^1$ Department of Physics, California State University, Fullerton, CA USA}

\affiliation{$^2$ Department of Computer Science, California State University, Fullerton, CA USA}

\altaffiliation{$^\star$  correspondence: wahmed@fullerton.edu}

\affiliation{$^3$ Department of Electrical Engineering, Princeton University, NJ, USA}
\affiliation{$^4$ Department of Physics, Massachusetts Institute of Technology, Cambridge, USA}


\date{\today}% It is always \today, today,

\begin{abstract}
Active matter systems exhibit rich emergent behavior due to constant injection and dissipation of energy at the level of individual agents.  Since these systems are far from equilibrium, their dynamics and energetics cannot be understood using the framework of equilibrium statistical mechanics.  Recent developments in stochastic thermodynamics extend classical concepts of work, heat, and energy dissipation to fluctuating non-equilibrium systems.  We use recent advances in experiment and theory to study the non-thermal dissipation of individual light-activated self-propelled colloidal particles.  We focus on characterizing the transition from thermal to non-thermal fluctuations and show that energy dissipation rates on the order of $\sim k_B T$/s are measurable from finite time series data.
\end{abstract}

\maketitle

%\tableofcontents

\section{Introduction}

Active colloids are self-propelled particles that convert chemical energy into directed mechanical motion at the microscopic-scale~\cite{bechinger2016active}.  They have become a paradigm in the active matter community because they exhibit emergent behavior~\cite{zottl2016emergent} such as phase transitions~\cite{cates2015motility} and dynamic crystallization~\cite{palacci2013living}, and are also the basis for studying non-equilibrium microscopic heat engines~\cite{martinez2017colloidal, pietzonka2019autonomous,ekeh2020thermodynamic, holubec2020active}. Significant effort has been put into developing a framework to understand active matter by connecting it to stochastic thermodynamics~\cite{seifert2012stochastic,speck2016stochastic,mandal2017entropy,szamel2019stochastic,fodor2016far}, which extends the concepts of classical thermodynamics to non-equilibrium systems and individual trajectories.  One general limitation of this approach is the entropy production cannot be fully inferred since the thermal and active noise cannot be explicitly disentangled along the trajectory~\cite{pietzonka2017entropy}.   Nevertheless, stochastic thermodynamics has potential to help move the field from studying specific phenomenological models of active matter to developing a general thermodynamic framework for driven active systems.  

Active matter systems are ubiquitous over a wide range of space and time scales~\cite{ramaswamy2010mechanics,marchetti2013hydrodynamics, gompper20202020}.  At the nanometer scale, single molecules can act as active matter~\cite{xu2019direct,jee2018catalytic}; at the microscopic scale, which is the most well-studied, biological and synthetic systems play the role of active matter~\cite{sanchez2012spontaneous,takatori2014swim,needleman2017active, prost2015active, duclos2017topological}; and, at the intermediate and larger scales, animals~\cite{attanasi2014information}, robots~\cite{scholz2018inertial}, human crowds~\cite{bottinelli2016emergent}, etc. operate as active matter. The underlying physical processes governing all of these systems varies widely e.g.~wet vs.~dry~\cite{marchetti2013hydrodynamics,chate2020dry}, under vs.~overdamped~\cite{takatori2015towards,sandoval2020pressure,klotsa2019above,lowen2020inertial}, thermal vs.~non-thermal~\cite{omar2020microscopic, van2019mechanical, cavagna2014bird}, etc.  However, they all have an important aspect in common --- non-equilibrium dynamics emerge because each individual element of the active matter system consumes energy and dissipates it via motion into the surrounding environment. This gives rise to emergent behavior that is not observable in equilibrium systems.  In an effort to understand this non-equilibrium behavior stochastic thermodynamics has emerged as a framework to quantify work, heat, and entropy fluctuations at the level of individual trajectories for non-equilibrium ensembles, leading to several predictions that can be tested experimentally~\cite{ciliberto2017experiments}. Furthermore, a flurry of activity has led to promising approaches to characterize non-equilibrium systems including broken detailed balance~\cite{gnesotto2018broken,mura2019mesoscopic,seara2018entropy, li2019quantifying}, Kullback-Leibler divergence~\cite{kullback1951information,roldan2012entropy,dabelow2019irreversibility}, information theory~\cite{parrondo2015thermodynamics,horowitz2014thermodynamics,martiniani2019quantifying}, breakdown of time-reversal symmetry~\cite{nardini2017entropy,maes2020frenesy, shankar2018hidden}, thermodynamic uncertainty relations~\cite{barato2015thermodynamic,horowitz2019thermodynamic,horowitz2017proof}, etc.  Our study focuses on quantifying dissipation of mechanical energy~\cite{harada2005equality,shinkai2014energetics}.

Towards connecting active matter and stochastic thermodynamics we use the simplest system, individual active colloids, and quantify their dynamics and energetics via experiments and simulations over a wide range of space, time, and activity.  We use light activated colloids, where activity is controlled via illumination intensity~\cite{palacci2014light}, observe them over a wide range of experimental timescales ($10^{-3}$ to $10^2$ s), and compare them to simulations of the Active Brownian Particle (ABP) model~\cite{romanczuk2012active}. By carefully tuning light-activation, our study focuses on the transition from thermal to non-thermal fluctuations of our colloidal system.  We apply relations from non-equilibrium statistical mechanics to characterize this transition in terms of forces and energy dissipation in the time- and frequency-domain~\cite{shinkai2014energetics,harada2005equality,ahmed2015active}.  Several theoretical relations for energy dissipation have been verified experimentally with optical traps or using numerical simulations~\cite{toyabe2007experimental,shinkai2014energetics,fodor2016nonequilibrium}, but here we apply them to a paradigmatic example of active matter where we can precisely tune activity --- light-activated colloidal particles~\cite{palacci2014light}.  Since we seek to extract non-equilibrium activity from finite time-series data, we focus on comparing our experiments to ABP simulations instead of analytic models.  Our results show that energy dissipation rates are measurable for even relatively low-activity colloids on the order of $\sim k_B T/$s. 

\section{Materials and methods}

\subsection{Preparation of colloids}
Light-activated colloids were prepared using previously published protocols~\cite{palacci2013living,palacci2014light}.  Briefly, hematite cubes were synthesized by dissolving 56 g of iron (III) chloride hexahydrate in 100 mL distilled water, followed by the addition of 90 mL of 6M NaOH while constantly stirring, and finally 10 mL deionized water.  The dissolved solution was baked in the oven for 8 days at 100 $^\circ$C.  Then the active colloids were created by embedding hematite cubes in TPM (Tripropylene glycol methyl ether) by  combining 2\% wt hematite solution in 98 mL deionized water, with 60 $\mu$L 28\% ammonia, and adding 1 mL of TPM while stirring at 400 rpm for 90 minutes, and then adding 1 mg AIBN (Azobisisobutyronitrile) and baking in the oven at 80 $^\circ$C for 3 hours. For active motion the colloids must be immersed in a ``fuel" buffer that supports the light-induced chemical reaction.  The fuel buffer used was composed of 65 $\mu$L of active colloids, 3.5 $\mu$L of TMAH (tetramethylammonium hydroxide), and 7.5 $\mu$L of 30\% hydrogen peroxide.  Our resulting active colloids have a radius of 1.5 to 2 $\mu$m.

\color{black}

\subsection{Microscopy and particle tracking}
Observation chambers were created using glass coverslips that were plasma-treated, and separated by double-stick tape to ``sandwich" the active colloid and fuel buffer.  A Nikon TE2000 with a 60x water-immersion objective (1.2 NA) and Hamamatsu ORCA-Flash4.0 V2 was used for microscopy.  Light activation was provided by an EXFO X-Cite 120PC through a 488 nm bandpass filter and light intensity measured at the sample plane varied from 0 (0\%), 3.3 (12\%), 5.7 (25\%), 11.9 (50\%), and 22.1 W/m$^2$ (100\%).  Brightfield image sequences were captured at sampling rates of $10^{3}$ to $10^0$ Hz.  Image-processing was done using FIJI~\cite{schindelin2012fiji} and MATLAB~\cite{MATLAB:2019a}, and single particles were tracked using polynomial fitting with Gaussian weight~\cite{rogers2007precise}.  Number of particles tracked for 0, 12, 25, 50, and 100\% experiments were 2296, 95, 118, 113, and 675 respectively. The maximum duration of recorded image sequences was 480 seconds, also corresponding to the maximum tracked trajectory length.  Typical trajectory lengths were shorter with an average length for 0, 12, 25, 50, and 100\% experiments of 91, 184, 188, 153, 89 seconds respectively.  No activity change due to depletion of fuel buffer was observed during the experimental time frame ($<$ 30 min).

\color{black}

\subsection{Numerical simulations}
The ABP model~\cite{romanczuk2012active} was used to simulate motion of active colloids in the overdamped regime (code published~\cite{volpe2014simulation}).  Parameters for simulation were extracted from analytic fits of experimental data.  Specifically two parameters are necessary to define simulation conditions:  the thermal diffusion coefficient ($D_\mathrm{th}$) and the average speed of self-propulsion ($v_0$).  From our experimental data we extract, $D_\mathrm{th} = 0.11$ $\mu$m$^2$/s, and for the five activity levels investigated, $v_0 =$ 0, 0.2, 0.36, 0.52, and 0.65 $\mu m/$s.  To explore short time-scale dynamics simulations with $\Delta t=10^{-3}$ s were run for $5 \times 10^5$ time steps for 100 particles.  For longer time-scale dynamics simulations with $\Delta t=10^{0}$ s were run for 500 time steps for 2000 particles.  Simulation parameters were chosen to be consistent with experimental conditions.

\begin{figure}[t]
\includegraphics[width=0.5\textwidth]{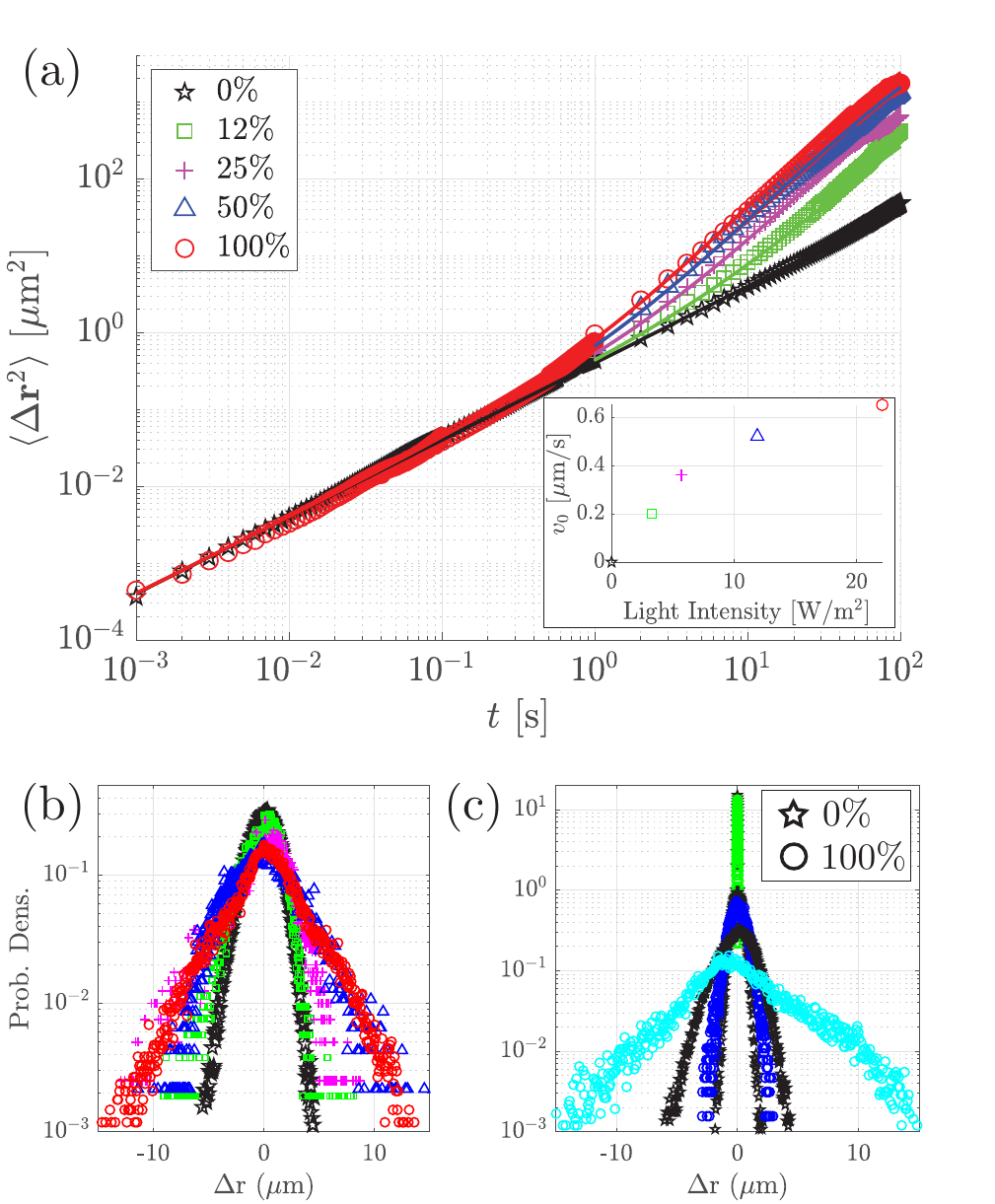}% Here is how to import EPS art
\caption{\label{fig:MSD} \textbf{Mean squared displacement and van Hove correlations for varying activity.} (a) In the absence of illumination (0\%), particles exhibit purely thermal diffusion ($\star$).  The average thermal diffusion coefficient for all particles was $D_{\mathrm{th}} \sim 0.11$ $\mu$m$^2$/s .  As light intensity is increased the MSD at longer timescales shows increased amplitude. SEM is smaller than symbol size and solid lines are theoretical fit to equation \ref{eqn:msd}.  Inset in (a) shows the average self-propulsion speed as a function of light intensity. (b,c) van Hove correlations are used to quantify the probability distribution of displacements. For increasing activity at $\Delta t = 10$ s, shown in (b), it is possible to observe a transition from Gaussian-like shape ($\star$) to non-Gaussian ($\circ$).  For varying timescales the effect is stronger, where thermal motion ($\star$) remains Gaussian and active motion ($\circ$) shows a much wider distribution with non-Gaussian tails as shown in (c) for $\Delta t = 10^{-2}, 10^0, 10^1$ s. } 
\end{figure}

\subsection{Data analysis for experiments and simulation}
All analysis of experiments and simulations was completed in MATLAB~\cite{MATLAB:2019a}.  The mean squared displacement (MSD) was calculated from positions, $\langle\Delta\mathbf{r}^{2}\rangle=\langle[r(t+\Delta t)-r(t)]^{2}\rangle$ with time and ensemble averaging.  A small number of particles ($<$ 1\%) appear stationary because they are stuck to the glass surface. Stationary particles were identified using MSDs near the noise limit and removed from further analysis. The van Hove correlations were used to characterize the probability distribution of displacements as done previously~\cite{ahmed2014active, hofling2013anomalous}, $P(\Delta x, \Delta t)$, where $\Delta x(\Delta t) = x(t + \Delta t) - x(t)$ and the distribution was normalized such that, $\int P(\Delta x,\Delta t)d\Delta x=1$. 

The power spectral density of a finite signal, $F(t)$ was estimated via, $PSD(F)=\frac{\tilde{F}\cdot\tilde{F}^{*}}{s\cdot p}$, where $\tilde{}$ is the fourier transform of the signal computed using the Fast Fourier Transform, $^*$ denotes the complex conjugate, $s$ is the sampling frequency, and $p$ is the number of data points in the time series.  The PSD was calculated for each individual trajectory and then ensemble averaged.  For calculating the average dissipation rate in the time domain, we first calculate the incremental dissipation, $\Delta \tilde Q(t)$, for each trajectory and average over all time to get a single value, $\Delta \tilde Q$; then we average over all trajectories to get $\langle \Delta \tilde Q / \Delta t \rangle$ where $\Delta t = 1$ second.

\section{Results and Discussion}

\subsection{Active diffusion and self-propulsion}
The overdamped dynamics of an active colloid can be conveniently described via the Langevin equation~\cite{sekimoto1998langevin,coffey2012langevin},
\begin{equation}
\gamma\dot{\mathbf{r}}=\mathbf{F}_{\mathrm{A}}+\sqrt{2 \gamma k_B T}\boldsymbol{\xi}
\label{eqn:eom}
\end{equation}
where $\mathbf{r}$ is position, $\mathbf{F}_\mathrm{A}$ is the active force, and $\sqrt{2\gamma k_B T}\boldsymbol{\xi}$ is the thermal force.  The thermal force is well-defined via the fluctuation-dissipation theorem in terms of a white noise term, $\boldsymbol{\xi}$, that satisfies $\langle\boldsymbol{\xi}(t)\rangle=0$ and $\langle\boldsymbol{\xi}(t)\boldsymbol{\xi}(t^{\prime})\rangle=\delta(t-t^{\prime})$, $T$ is temperature, $k_B$ is the Boltzmann constant, and $\gamma = 6 \pi R \eta$ is the friction coefficient where $\eta$ is the fluid viscosity~\cite{kubo1966fluctuation}. The active force, $\mathbf{F}_\mathrm{A}$, is more interesting because it is the source of all non-equilibrium dynamics in the system.  The most common form of the active force for self-propelled particles is $\mathbf{F}_\mathrm{A} = \gamma \mathbf{u}$ where $\mathbf{u}$ is a stochastic velocity with statistics determined by the underlying model, e.g. run-and-tumble, active Ornstein-Uhlenbeck, and active Brownian~\cite{fodor2018statistical}. These models differ in their higher order statistical properties~\cite{shee2020active} but have an identical MSD.

Perhaps the most widely used approach to quantify active diffusion of colloids is to simply examine the MSD and how much it deviates from thermal motion of the same colloid \cite{fodor2018statistical},
\begin{equation}
\langle\Delta\mathbf{r}^{2}(t)\rangle=4(D_{\mathrm{th}}+D_{\mathrm{ac}})\vert t\vert+2(v_{0}\tau)^{2}\left(\mathrm{e}^{-\vert t\vert/\tau}-1\right)
\label{eqn:msd}
\end{equation}
where $D_\mathrm{ac} = v_0^2 \tau /2$ is the active diffusion coefficient, $v_0 = \vert \mathbf{u} \vert$ is the self-propulsion velocity, and $\tau$ is the persistence time.  Two limiting cases emerge where at short timescales ($t \ll \tau$) thermal diffusion dominates, $\langle\Delta\mathbf{r}^{2}(t)\rangle = 4 D_{\mathrm{th}}\vert t\vert$, and at long timescales ($t \gg \tau$) we have active diffusion which is a combination of thermal and active processes, $\langle\Delta\mathbf{r}^{2}(t)\rangle = 4(D_{\mathrm{th}}+D_{\mathrm{ac}})\vert t\vert$.  If a wide enough range of timescales is observed, then it is possible to extract both $D_{\mathrm{th}}$ and $D_{\mathrm{ac}}$ by fitting the MSD.

We use equation \ref{eqn:msd} to characterize the transition from thermal to actively driven diffusion of our active colloids.  Experimental data and theoretical fits are shown in Fig. \ref{fig:MSD}a.  Here we see that thermal diffusion dominates at shorter timescales ($t<1$ s).  At longer timescales we observe the transition to active motion with increasing illumination intensity. Self-propulsion velocities extracted from fits range from $v_0 \sim 200 - 650$ nm/s (see inset, Fig. \ref{fig:MSD}a), indicating even small levels of activity are visible in the MSD; and self-propulsion velocity increases sub-linearly with illumination intensity as expected for the diffusiophoresis process~\cite{palacci2014light}.   Timescale of the persistence is $\tau \sim 48$ s, which is consistent with the rotational diffusion time, $\tau_{r}=8\pi\eta R^{3}/k_{B}T$, for a 2 $\mu$m particle.  This results in active diffusion coefficients, $D_\mathrm{ac} \sim 0.9 - 6.2$ $\mu$m$^2$/s roughly an order of magnitude larger than for passive diffusion ($D_\mathrm{th} \sim 0.11$ $\mu$m$^2$/s).  Interestingly, the $D_\mathrm{ac}$ for our active colloids is similar to that observed in living cells~\cite{colin2020active}. 

To further investigate displacement fluctuations we use van Hove correlations to quantify the probability distribution of displacements for a given activity and timescale (Fig.~\ref{fig:MSD}b,c).  When activity is increased the distribution at timescale $\Delta t = 10$ s gradually broadens and becomes more non-Gaussian as expected (Fig.~\ref{fig:MSD}b).  The effect is even more pronounced for the highest activity level as timescale is increased (Fig.~\ref{fig:MSD}c).  Non-Gaussian distributions indicate non-equilibrium behavior dominates at longer timescale.  This type of displacement fluctuation analysis is commonly used in quantifying microscopic motion that is a mixture of thermal and non-thermal fluctuations~\cite{toyota2011non,stuhrmann2012nonequilibrium,ahmed2014active}, but in the following sections we extend our analysis to more recently developed approaches related to forces and dissipation.

\color{black}
\begin{figure}[t]
\includegraphics[width=0.5\textwidth]{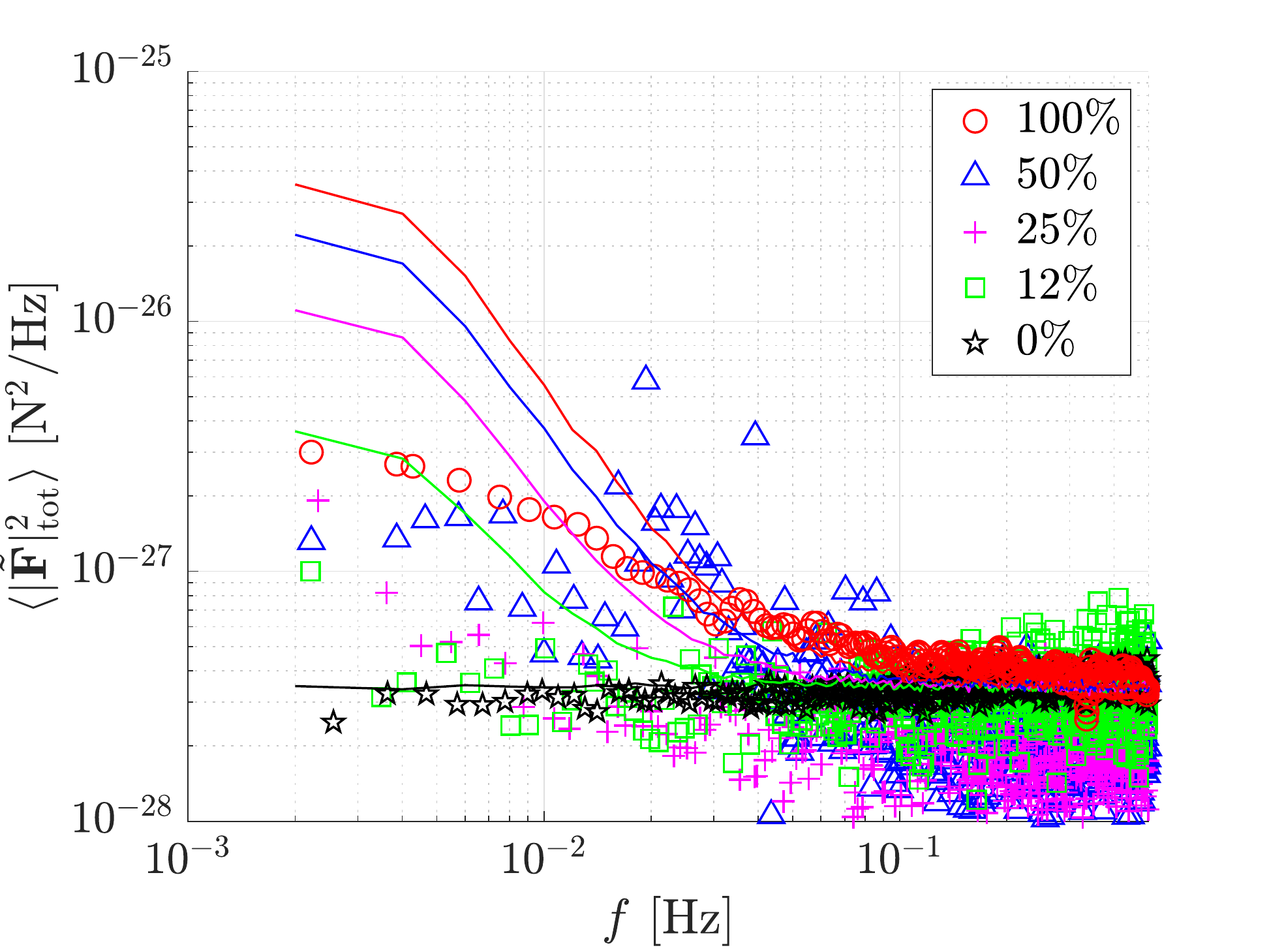}
\caption{\label{fig:PSDForce} \textbf{The total force spectrum of an active colloid.}  Results for both experiment (symbols) and simulation (lines) are shown for all activity levels.  Force spectra are much larger at lower frequencies due to non-equilibrium activity while higher frequency forces are mainly thermal.  Simulations show a clear incremental increase in the force spectrum with activity (averaged over 2000 particles). Experiments show a similar trend with greater noise.}
\end{figure}

\subsection{The force spectrum}
A more recently developed approach to quantify microscopic non-equilibrium dynamics is the force spectrum~\cite{gallet2009power,guo2014probing,ahmed2018active,bohec2019distribution}, which is the power spectrum of the stochastic forces experienced by the particle.  The force spectrum contains more information than the MSD, because it incorporates the mechanical properties of the system.  In the case of an active colloid in an aqueous solution, the total force spectrum can be estimated from the stochastic trajectories assuming low Reynolds Stokes flow~\cite{happel2012low}.  From each trajectory we calculate the total force as $F_\mathrm{tot}(t) = \gamma \vert \dot{\mathbf{r}}(t) \vert$ where $\vert \dot{\mathbf{r}}(t)\vert = \vert \Delta \mathbf{r}(t) \vert / \Delta t$ and $\Delta$ is the incremental change between two consecutive frames.  Strictly speaking this velocity is not well-defined for a stochastic process, and in-depth discussion of the statistical considerations of time resolution can be found elsewhere~\cite{shinkai2014energetics, sekimoto2010stochastic}.

$F_\mathrm{tot}$ is the instantaneous force experienced by the colloid along its trajectory and thus we can estimate the force spectrum by calculating its power spectral density.  The force spectrum for experimental data and simulations is shown in Fig.~\ref{fig:PSDForce}.  Here we see at high frequencies thermal forces are dominant, and at lower frequencies non-equilibrium forces become apparent and are proportional to light intensity.  We find qualitative agreement between experimental and simulated force spectrum, but in general the force spectra from experiments are lower than simulation at low frequencies.  This consistent difference arises because highly active colloids have shorter duration trajectories and fewer statistics in experiments simply because they leave the field of view.  Therefore, low-frequency (long timescale) activity is often under-estimated in experimental measurements.  We note that 0\% and 100\% data have a larger number of particles and thus have less experimental noise, however in all cases experiments are biased towards observing ``lower activity'' colloids because of particles leaving our limited field of view.  This effect has been used to direct self-assembly of structures~\cite{arlt2018painting}.  Nevertheless, the force spectrum of the active colloids is clearly distinguishable from the non-active colloids.  This highlights the sensitivity of the force spectrum, because in the time domain, the average force expected due to self-propulsion for our highest level of activity is, $\langle F\rangle=6\pi R\eta v_{0}\approx25$ fN, which is vanishingly small and within the measurement noise.  However, when quantifying the stochastic forces using the force spectrum the non-equilibrium activity is clearly evident.  For comparison, forces are one to two orders of magnitude larger for micro-swimmers~\cite{elgeti2015physics} and organelles/particles of comparable size in cells~\cite{gallet2009power,guo2014probing, ahmed2018active} --- usually at the pN scale.

\color{black}

\begin{figure}
\includegraphics[width=0.5\textwidth]{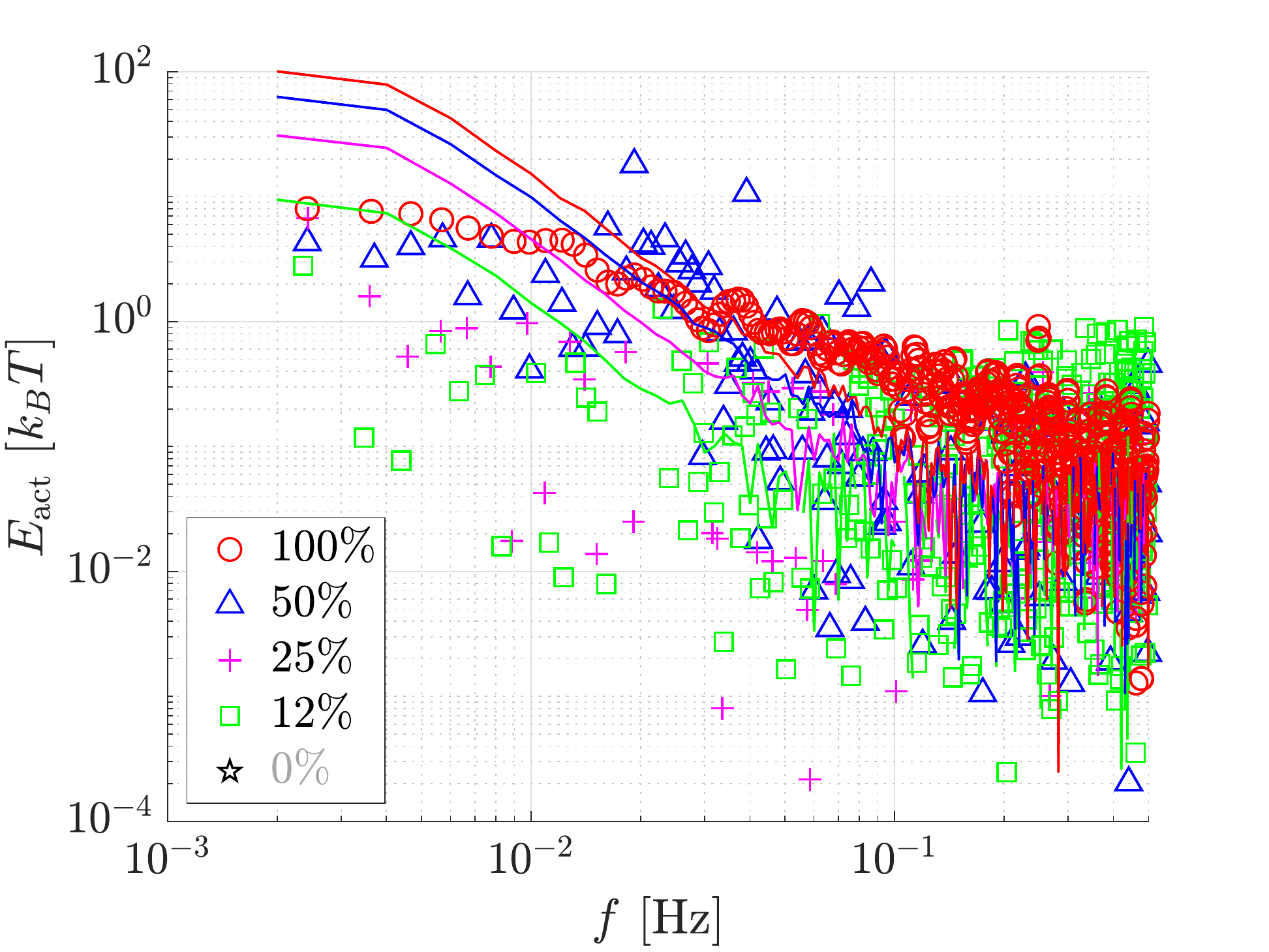}% Here is how to import EPS art
\caption{\label{fig:Eactive} \textbf{Isolating non-equilibrium activity.} The active energy quantifies the amplitude of non-equilibrium fluctuations as a function of frequency in units of thermal energy. The general trend persists where increased activity results in higher active energy.  The active energy is directly related to the dissipation spectrum, $\tilde{I}$, from the Harada-Sasa equality and can be integrated to get the average energy dissipation rate.}
\end{figure}

\subsection{Activity in the frequency domain}
To characterize the activity of the colloid we must isolate the non-equilibrium behavior by removing the thermal fluctuations. This can be done in several ways, e.g. the force spectrum~\cite{ahmed2015active}, the effective energy~\cite{cugliandolo2011effective}, and the Harada-Sasa equality~\cite{harada2005equality}, which all give essentially the same information.  To calculate the active force spectrum our first assumption is that the total force that drives motion of the colloid is the sum of the active and thermal forces, $\mathbf{F}_\mathrm{tot} = \mathbf{F}_\mathrm{A} + \mathbf{F}_\mathrm{th}$.  Moving to the frequency domain and calculating the power spectrum, we have $\langle \vert \tilde{\mathbf{F}}_\mathrm{tot} \vert ^2 \rangle=  \langle \vert \tilde{\mathbf{F}}_\mathrm{A} \vert^2 \rangle +  \langle \vert \tilde{\mathbf{F}}_\mathrm{th} \vert^2 \rangle + 2\langle \vert \tilde{\mathbf{F}}_\mathrm{A} \tilde{\mathbf{F}}_\mathrm{th} \vert \rangle$
where for synthetic self-propelled particles we can assume that the active force and the thermal force are independent (exhibit no time correlations), and thus the cross-term vanishes allowing us to disentangle active and thermal forces.  The result is we can estimate the active force spectrum, by simply subtracting the thermal spectrum from the total spectrum, 
\begin{equation}
\langle \vert \tilde{\mathbf{F}}_\mathrm{A} \vert^2 \rangle = \omega^2 \gamma^2 \tilde{C} - 2 \gamma k_\mathrm{B}T
\label{eqn:F_A}
\end{equation}
where $\omega$ is frequency in rad/s and $\tilde{C}$ is the power spectral density of position.  The active force spectrum, $\langle \vert \tilde{\mathbf{F}}_\mathrm{A} \vert^2 \rangle$, quantifies the stochastic forces driving the colloid motion that originate from only non-equilibrium sources.

A related approach is to calculate the effective energy~\cite{cugliandolo2011effective}, where we start with the familiar relation based on violation of the fluctuation-dissipation theorem, $E_{\mathrm{eff}}=k_{B}T_{\mathrm{eff}}=\omega\tilde{C}/2\tilde{\chi}^{\prime\prime}$, where $\tilde{\chi}^{\prime\prime}$ is the imaginary component of the response function~\cite{fodor2016nonequilibrium}.  The effective energy can be re-written in terms of the force spectrum, 
$E_{\mathrm{eff}}=\langle\vert\mathbf{F}_{\mathrm{tot}}\vert^{2}\rangle/\langle\vert\mathbf{F}_{\mathrm{th}}\vert^{2}\rangle$, in units of $k_B T$.  Thus the thermal energy can be subtracted from $E_\mathrm{eff}$ yielding the active energy,
\begin{equation}
E_\mathrm{act} = \frac{\omega^{2}\gamma\tilde{C}}{2k_{B}T}-1
\label{eqn:E_A}
\end{equation}
shown for experiments and simulations in Fig.~\ref{fig:Eactive}.  The active energy, $E_\mathrm{act}$, is a frequency dependent measure of the amplitude of energetic fluctuations driven by non-equilibrium sources.  Here we see at low frequencies the active energy is on the scale of 1-100 $k_B T$ and becomes vanishingly small at higher frequencies ($f > 10^{-1}$ Hz). The amplitude of our energetic fluctuations is the same order of magnitude as quantified in other non-equilibrium systems at the microscopic-scale such as red blood cells~\cite{betz2009atp,ben2011effective,turlier2016equilibrium}, mammalian cells~\cite{wilhelm2008out,gallet2009power,ahmed2018active}, and colloidal glasses~\cite{greinert2006measurement,jabbari2007fluctuation,bellon2001violation}; however we observe this activity at lower frequencies ($f < 10^{-2}$ Hz).  

While the effective/active energy is a quantitative metric for how far from thermal equilibrium a system is, it is not an energy in the traditional sense because it is a function of frequency.  Perhaps a more meaningful metric of non-equilibrium activity is the average energy dissipation rate due to only non-equilibrium processes.  To calculate this we use the Harada-Sasa equality where we can estimate the dissipation spectrum from experimentally measurable quantities~\cite{harada2005equality,toyabe2007experimental},
\begin{equation}
\tilde{I}=\omega\gamma\left(\omega\tilde{C}+2k_{\mathrm{B}}T\tilde{\chi}^{\prime\prime}\right)
\label{eqn:Idiss}
\end{equation}
where we can see in thermal equilibrium, $\tilde{I} = 0$ due to the fluctuation-dissipation theorem~\cite{kubo1966fluctuation}. The dissipation spectrum, $\tilde{I} (\omega)$, is equivalent to the active energy shown in Fig.~\ref{fig:Eactive}.  Thus for non-equilibrium steady states the energy dissipation rate can be estimated via, $\langle J\rangle=\int d\omega\tilde{I}(\omega)/2\pi$, as discussed later. These three approaches for quantifying non-equilibrium activity involve the force spectra, which required ensemble averaging due to the finite length of our time domain signals.  This is generally true for estimating power spectrum from short trajectory data sets~\cite{krapf2018power,krapf2019spectral}.  Thus using these frequency-domain approaches with short trajectories (less than 500 time steps), we can only isolate the non-equilibrium activity on average and not of individual stochastic trajectories.

\color{black}

\begin{figure}[h]
\includegraphics[width=.45\textwidth]{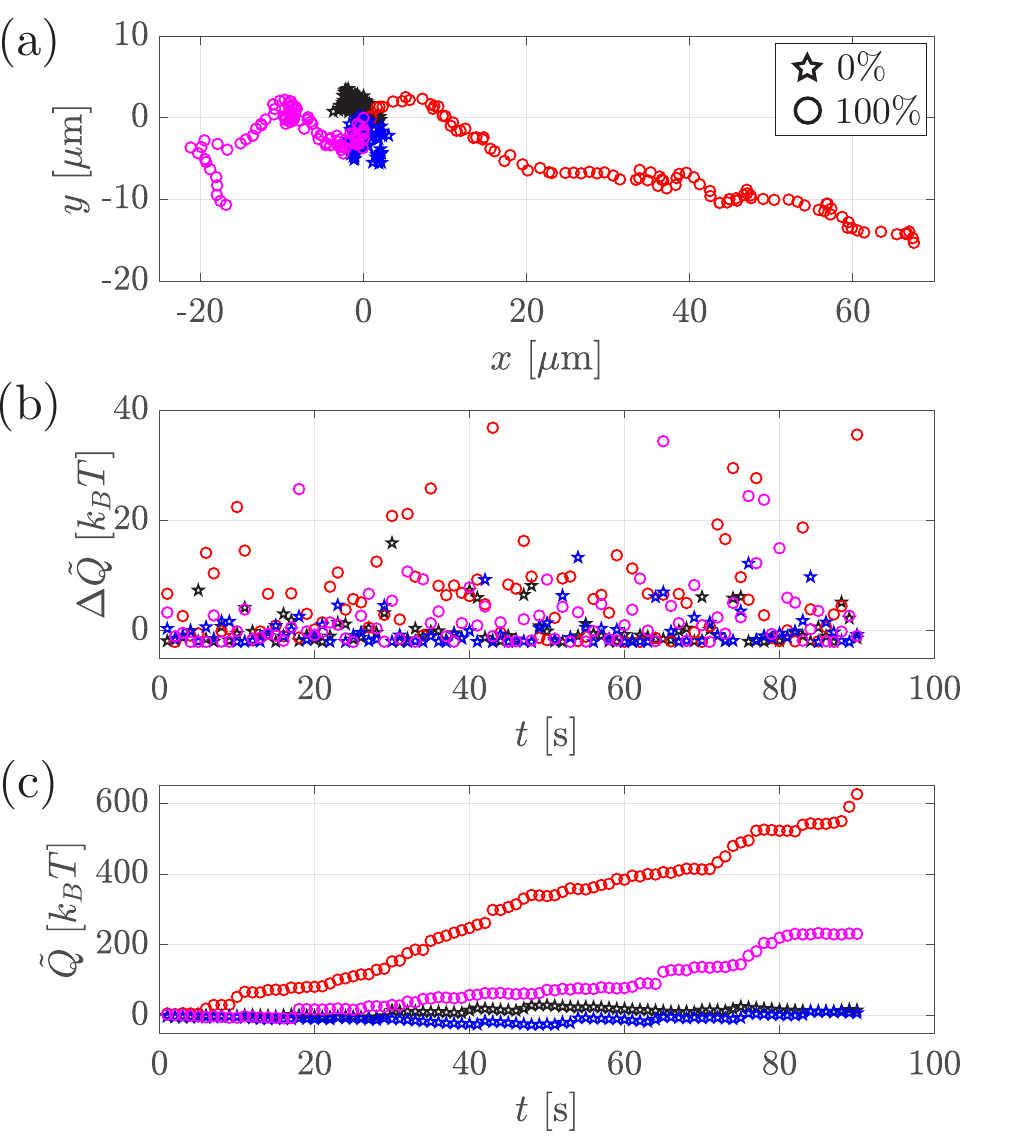}% Here is how to import EPS art
\caption{\label{fig:dQ} \textbf{Activity in the time domain.} (a) In the absence of light-activation ($\star$) representative colloids exhibit very small motions due to only thermal fluctuations, whereas light-activated colloids ($\circ$) clearly show persistent motion over tens of microns.  (b) Active colloids ($\circ$) show larger $\Delta \tilde{Q}$ than non-active ($\star)$. (c) When integrated, the accumulated dissipation, $\tilde{Q}$ shows a linear increase in non-thermal dissipation with time for active colloids, but no growth for non-active colloids which exhibit only thermal fluctuations.}
\end{figure}

\begin{figure}[h!]
\includegraphics[width=.45\textwidth]{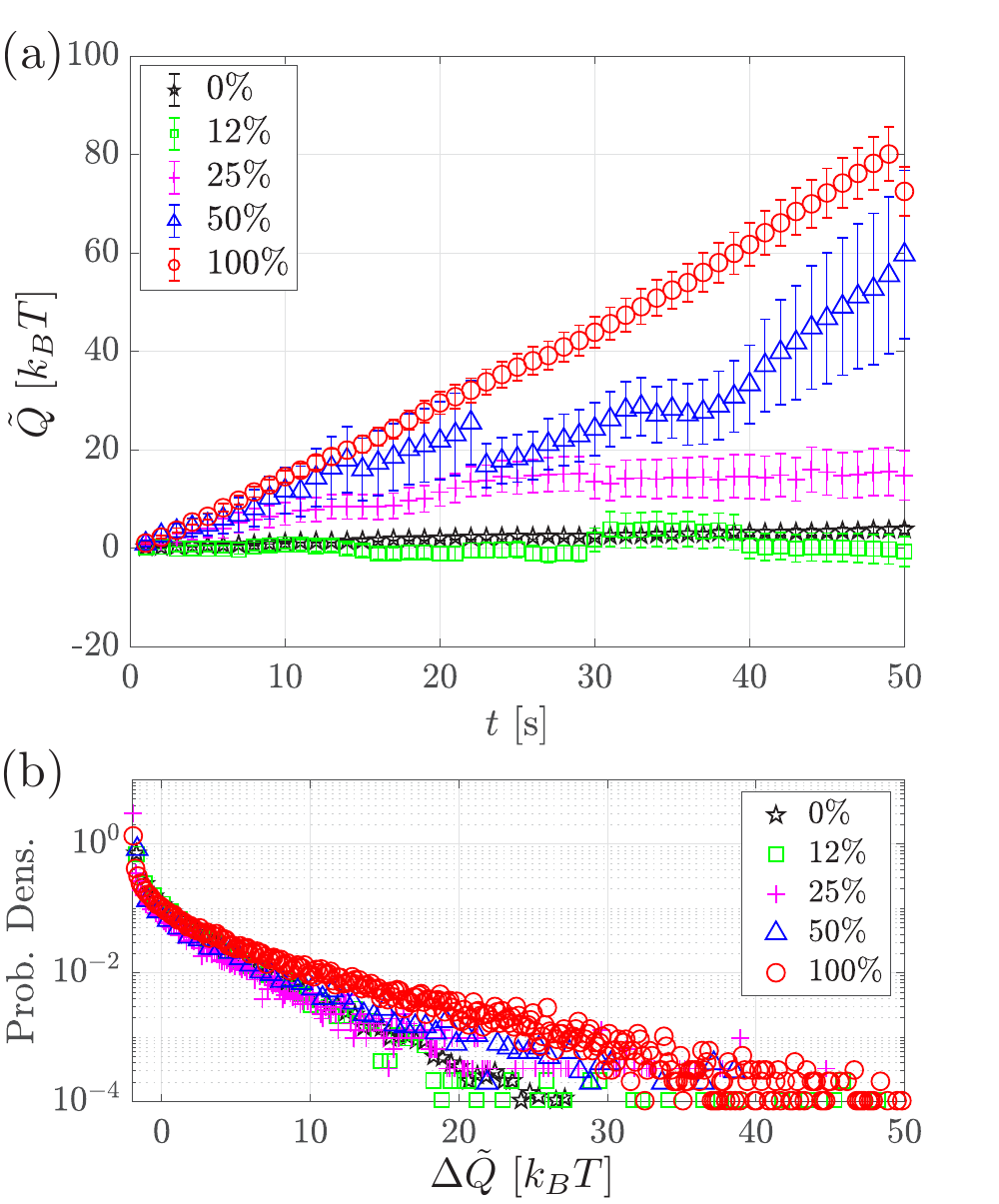}% Here is how to import EPS art
\caption{\label{fig:Q} \textbf{Average dissipation with time.} (a) In the absence of light-activation (0\%) colloids on average exhibit no excess heat dissipation ($\star$).  As illumination intensity increases, the excess dissipation becomes more evident -- especially at longer times. Error bars are SEM. (b) The probability density of heat fluctuations shows a longer tail with increasing activity, which is most evident for the highest activity ($\circ$).}
\end{figure}

\subsection{Activity in the time domain}
To characterize the energetics of individual stochastic trajectories we use an approximation of the heat dissipation in the time domain $(\Delta \tilde{Q})$, which has been proposed as an alternative to MSD analysis to quantify non-equilibrium systems~\cite{shinkai2014energetics},
\begin{equation}
\Delta \tilde{Q}(t)=\gamma\frac{(\Delta x_{t})^{2}}{\Delta t}-2k_{\mathrm{B}}T
\label{eqn:dQ}
\end{equation}
where $\Delta x_t$ is the displacement between two consecutive positions in a trajectory separated by time $\Delta t$.  It is important to note that $\Delta \tilde{Q}$ is an estimate of the excess energy dissipated after the equilibrium dissipation is subtracted; for details see the derivation by Shinkai and Togashi~\cite{shinkai2014energetics}.  Equation \ref{eqn:dQ} is calculated without time/ensemble average and thus estimates the dissipation along a single trajectory as shown in Fig.~\ref{fig:dQ}.  Here in Fig.~\ref{fig:dQ}a we see at the trajectory level, active particles ($\circ$) are clearly distinguishable from thermal fluctuations ($\star$). The incremental dissipation, $\Delta \tilde{Q}$, shows a much larger range of dissipation for active colloids (Fig.~\ref{fig:dQ}b).  When integrated, the accumulated dissipation of a trajectory, $\tilde{Q}$, shows active particles ($\circ$) exhibiting dissipation that grows roughly linear with time consistent with the ABP model with constant propulsion (Fig.~\ref{fig:dQ}c).  Conversely, passive particles ($\star$) exhibit dissipation that fluctuates around zero over time.

By averaging over ensembles (but not time) we can analyze the average dissipation with time as shown in Fig.~\ref{fig:Q}a.  Here we see that the more active colloids (50, 100\%) begin to show significantly higher accumulated dissipation  at $t>10$ s, which is consistent with activity emerging in the force spectrum and active energy at $f<10^{-1}$ Hz.  Looking at the probability density of heat fluctuations (Fig.~\ref{fig:Q}b) we can see that the activity skews the distributions to have longer tails with increasing activity. This shift is expected for small highly fluctuating and active systems~\cite{bustamante2005nonequilibrium}.

\subsection{Average dissipation rate}
\begin{figure}[h]
\includegraphics[width=.5\textwidth]{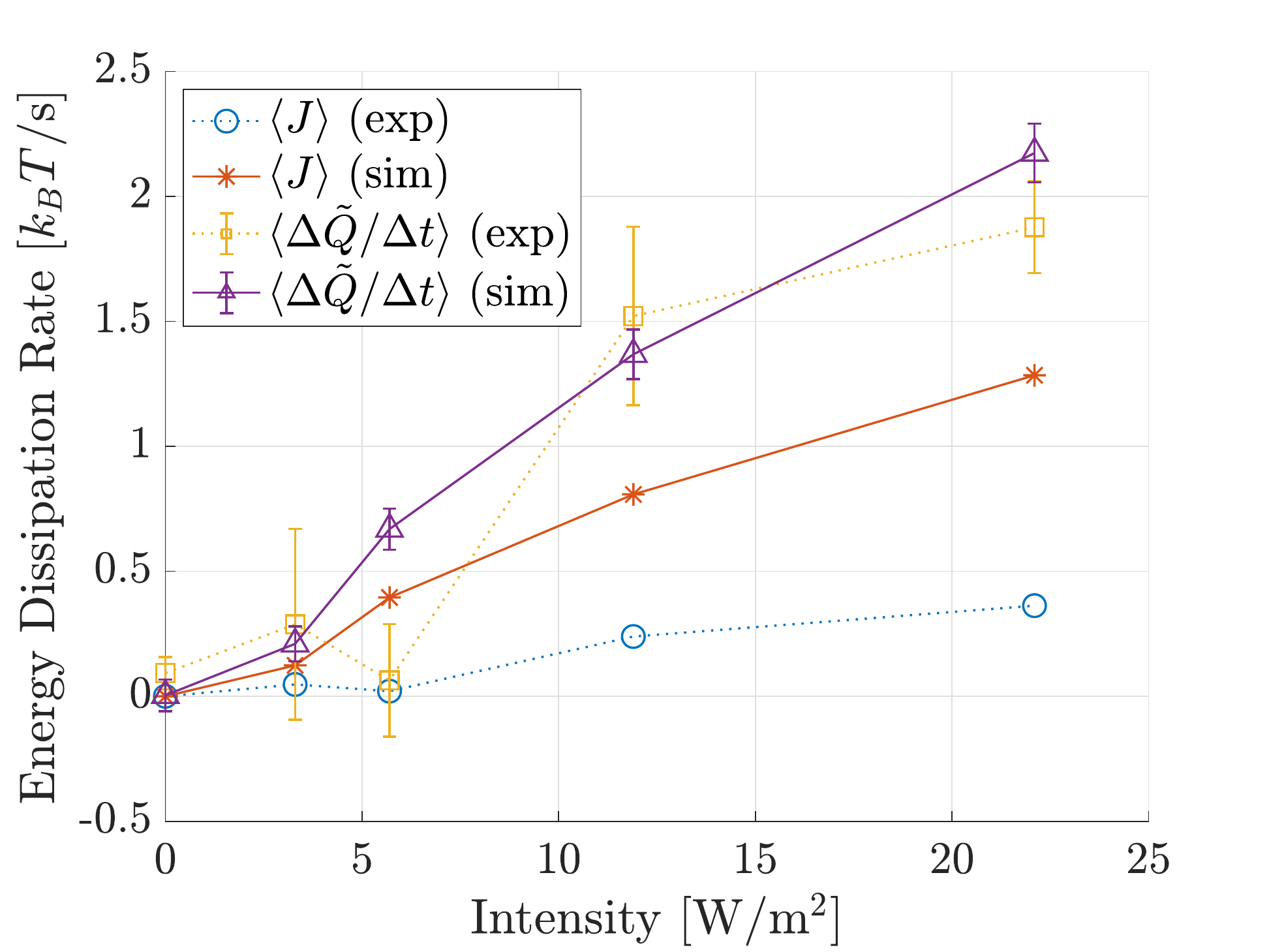}% Here is how to import EPS art
\caption{\label{fig:J} \textbf{Average energy dissipation rate.} Here we compare the energy dissipation rate calculated for experiments (exp) and simulation (sim) using two different methods --- $\langle J \rangle$ from the Harada-Sasa equality~\cite{harada2005equality} and $\langle \Delta \tilde{Q}/ \Delta t \rangle$ calculated from the Shinkai-Togashi expression~\cite{shinkai2014energetics}.}
\end{figure}

We applied two different approaches for calculating energy dissipation due to non-equilibrium processes for our active colloid experiments and simulations: (1) The Harada-Sasa equality to calculate the ensemble averaged energy dissipation spectrum in the frequency-domain, $\tilde{I}(\omega)$, and numerically integrated over the experimentally accessible frequencies to calculate the average energy dissipation rate, $\langle J \rangle$~\cite{harada2005equality}; and (2) The Shinkai-Togashi expression to calculate the incremental energy dissipation as a function of time for individual trajectories, $\Delta \tilde{Q}$, and subsequently applied time and ensemble averaging to calculate the average energy dissipation rate, $\langle \Delta \tilde{Q} / \Delta t \rangle$~\cite{shinkai2014energetics} (see methods for details).  In principle, in the limit of $t \rightarrow \infty$ (infinite time series data), these two measures of non-equilibrium activity should converge to the same value, $\langle J \rangle \approx \langle \Delta \tilde{Q} / \Delta t \rangle$.  As shown in Fig.~\ref{fig:J}, the two metrics, $\langle J \rangle$ and $\langle \Delta \tilde{Q} / \Delta t \rangle$  agree qualitatively, but their numerical values differ with the $\langle J \rangle$ being consistently lower. However, given that this work focuses on the transition from thermal to non-thermal fluctuations (i.e. low levels of activity) for finite time series data, these two metrics are surprisingly consistent --- showing that non-equilibrium dissipation becomes significant at activity levels corresponding to $v_0 > 0.5$ $\mu$m/s (light activation intensity $> 11.9$ W/m$^2$) and average rates on the order of $\sim k_B T/$s.  This is quantitatively consistent with the classical estimate of the power required to drive a micron-sized probe through water.  The energy dissipation rates measured here are significantly smaller than other systems such as single molecules ($\sim 20$ $k_B T$/s~\cite{toyota2011non}), interacting driven colloids ($\sim 200$ $k_B T$/s~\cite{lander2012noninvasive}), and organelles in cells ($\sim 360$ $k_B T$/s~\cite{fodor2016nonequilibrium}), highlighting the sensitivity of this approach. These results show that with time and ensemble averaging, even low-levels of non-equilibrium activity can be extracted using both the Harada-Sasa equality in the frequency-domain and the Shinkai-Togashi expression in the time-domain.

\section{Conclusion}

This work applies recently developed relations from stochastic thermodynamics~\cite{harada2005equality,shinkai2014energetics} to investigate the transition from equilibrium to non-equilibrium dynamics of light-activated self-propelled colloids via experiments and simulations. Our results show that energy dissipation rates on the order of $\sim k_B T/$s are measurable for individual colloids from finite time series data. The activity of our individual active colloids is orders of magnitude smaller than comparable scale transport in living cells, yet still measurable. This helps set the stage to move beyond displacement fluctuation analysis --- and to quantify active particles in terms of forces and energetics towards the development of non-equilibrium thermodynamics of active matter.

\section{Conflicts of interest}
There are no conflicts of interest to declare.

\section{Acknowledgements}
The authors acknowledge fruitful discussions with \'{E}. Fodor (Univ.~of Cambridge) and J. Palacci (UC San Diego), an initial gift of self-propelled colloids from M. Driscoll (Northwestern University), and guidance on colloidal synthesis from the Sacanna Lab (New York University).  WWA acknowledges financial support from the Cal State Fullerton Research Scholarly and Creative Activity Grant. SE, MG, PB-P acknowledge the Dan Black Family Fellowship. NL acknowledges CSUF Project RAISE, U.S. Department of Education HSI-STEM award number P031C160152.

\bibliography{main_Colloid.bbl}% Produces the bibliography via BibTeX.

\end{document}